# Understanding plasticity in zirconium using in-situ measurement of lattice rotations

Vivian Tong[1,2], Euan Wielewski[3], Ben Britton*[2]

1. National Physical Laboratory, Teddington
2. Imperial College London, Prince Consort Road, London, SW7 2AZ
3. Anomalous.ai

## Abstract

Understanding deformation in polycrystalline metals is critical to use them in high-value high-risk applications. We present in-situ characterisation of plastic deformation of zirconium, a hexagonal closed packed (HCP), metal. Analysis of plastic deformation is performed using electron backscatter diffraction (EBSD) to reveal changes in lattice orientation. Through implementation of TrueEBSD, we can relate the lattice rotations back to the undeformed reference frame. This enables us to explore which slip systems are active and the degree of homogeneous (i.e. deformation with respect to the external load) and heterogeneous (i.e. deformation with respect to the local grain neighbourhood). Additionally, from our analysis, we notice that lattice rotations consistent with a significant fraction <a> pyramidal slip are found. These results are placed in the context of deformation and performance of HCP alloys and zirconium alloys used as nuclear fuel cladding.

## Graphical Abstract

## Introduction

Understanding the deformation of materials is critical for their implementation in high-value high-risk applications, where the lifetime of safety critical components is important for economic and societal benefit. To develop our understanding, we must develop characterisation approaches and techniques that enable us to 'see' the mechanisms that carry strain, such as plastic slip through dislocation motion.

In polycrystalline materials, these mechanisms typically operate over a number of length scales. At the nanometre length scale, plastic slip is carried by motion of individual dislocations and their cooperative motion with response to the (evolving) local boundary conditions. The collective motion of dislocations at the micrometre length scale can typically be observed throughs slip trace analysis, the change in shape of the structure through image based digital image correlation.

Historically analysis of these deformation modes involves 'snap shot' imaging using transmission electron microscopy, where the Burgers vector and line direction can be relatively easily determined through dark field and/or weak beam imaging modes, but in this case the slip plane is presumed. While edge dislocation segments may be present, is remains unclear if those dislocations are the ones which enabled plastic slip. Furthermore, it is even less clear for screw dislocations as their active slip plane is necessarily unknown, as they may cross slip as needed. Detailed and careful *in-situ* TEM observations [1] shed light on critical aspects of the operating slip modes, however the boundary conditions of these experiments is typically limited due to the thin films. One further complication is that mobile and sessile dislocations contribute in different ways to plasticity and work hardening, but these cannot be deterministically separated by *ex-situ* observations.

More recent developments to understand plastic slip include use of so-called "nano-DIC" of "high resolution DIC" where digital image correlation (DIC) is performed using nm sized speckle patterns [2, 3]. Deformation is tracked by sub-windowing the field of view and performing image registration of each sub-window between captured timesteps to create a two-dimensional field of displacement vectors. The displacement field can be used to calculate the (in-plane) total strain field and evaluated to explore deformation localised at slip bands [3]. While the 2D DIC-based slip trace analysis is clearly of interest, there are several complications. Formally, in one limit, the extent and patterning of strain localisation is dependent on the size and spacing of the sub-windows. Next, DIC-based measurements can identify the surface trace of an active slip plane and in-plane slip direction along the slip trace, but not the dislocation type(s) which contributed to the localised strain. DIC-based slip system analysis becomes ambiguous if multiple easy slip directions share a crystallographic slip plane, or multiple easy slip planes intersect the sample surface with similar plane traces. This ambiguity can be reduced if the displacement vector is considered such as in the relative displacement ratio (RDR) [4], i.e. the in-plane displacements along a slip trace, but the possible crystallographic slip directions are assumed *a priori*. Recent work by Shi et al. [5]has indicated that out of place displacements are accessible with SEM based stereoscopic methods using surface markers, but the out of plane resolution (and analysis out of plane slip traces is not accessible due to spatial resolution limits). Applying this method, together with HR-DIC, could be interesting.

For deformation of hexagonal close packed metals, there are a limited number of slip systems that operate. In zirconium, the major slip planes are basal, prismatic, and the $1^{st}$ and $2^{nd}$ order pyramidal planes [6]. The easiest method for the crystal to deform is via slip with <a> type dislocations, on the basal, prismatic and potentially $1^{st}$ order pyramidal planes. <a> type slip does not enable strain along the <c> axis, and the crystal can deform with <c+a> type slip on the $1^{st}$ order or $2^{nd}$ order pyramidal planes, or T1 type extension twins may operate. Gong et al. [7] have directly fabricated cantilevers oriented to activate <a> basal, <a> prism, and <c+a> pyramidal slip systems and used the measured critical resolved shear stress (CRSS). *Ex-situ* analysis of TEM data reveals the presence of <a> type dislocations, and on occasions pile-ups of <a> dislocations on prismatic or basal planes, but there is necessary uncertainty about the slip plane of <a> screw dislocations.

Electron backscatter diffraction [8] is an alternative method of analysing the impact of deformation on the crystal lattice. In a constrained volume of polycrystals, increasing plastic slip can be correlated to increasing local rotation gradients, either through Kernel Average Misorientation (KAM), which is the misorientation of one point with respect to its neighbourhood [9]; the grain orientation spread (GOS), which is the average change of grain orientation within a grain [10]; or more formalised assessment of the lattice curvature through Nye's analysis [11]. We note with care here that these lattice curvature assessments only explore the residual dislocation 'debris' left over after deformation has occurred and may not reflect the active slip systems. Wright et al. [10] have observed in a deforming a polycrystalline S23C medium carbon steel with uniaxial tension that the sample averaged GOS and KAM values tend to increase linearly with increasing strain.

We can imagine the build-up of dislocations near grain boundaries, due to blocking of the slip system at the interface, and the build-up of local curvature from "excess dislocations", due to the operation of multiple slip systems which results in tangles that can be observed as which create lattice curvature dependant on the dislocation structures observed and the step size [12]. These approaches have proven insightful, but typically due to ambiguity of the curvature to GND density calculations, the activation of specific slip systems is rarely considered (except in a few examples where a specific slip system has operated, e.g. associated with a clear slip band [13, 14] and slip activation in the neighbour grain, or in crystals where there are a limited number of slip systems operating such as in minerals [15]).

An alternative approach in understanding slip in polycrystalline materials is to track the evolution of grain-scale lattice rotations [16-18]. Slip induces lattice rotation in polycrystal deformation. Dislocation motion in a grain causes 'deck of cards'-like sliding of the crystal either side of the slip plane, which changes the shape of the grain in an unconstrained material. In a grain constrained by its local neighbourhood, such as grain boundaries or the remainder of the grain, the shape change (plastic rotation) from dislocation slip must be compensated by an equivalent crystal rotation (i.e. the lattice spin), where the axis of rotation is given by the dominant lattice rotation, which is also called the Taylor rotation [19, 20] and the axis can be calculated from: $\boldsymbol{\omega} = \boldsymbol{b} \times \boldsymbol{n}$, where $\boldsymbol{b}$ is the Burgers vector and $\boldsymbol{n}$ is the slip plane normal. Here we note that the lattice spin is different to the plastic spin, and we are not exploring the difference in our analysis. In cases where there is limited constraint, the plastic spin may dominate [21].

If the Taylor axis is represented in the crystal reference, for single slip in zirconium (a hexagonal closed packed crystal) the analysis of the Taylor rotation axis can be illuminating as the axis for the different slip systems are reasonably well spaced and can be plotted, see Figure 1, using a colourmap based upon Inversed Pole Figure colouring (as often used for EBSD measurements of grain orientation). This is similar to the analysis provided by Wright et al. [10] for cubic samples, but Wright et al. only commented on spatial variations in the Grain Rotation Orientation Deviation (GROD) axis rather than ascribing them to particular slip systems as we will do later. Analysis of the total rotation about this axis is broadly captured within the so-called GROD maps, however as Wright et al. note the frame in which this GROD is considered (i.e. with respect to the grain mean orientation or an orientation deviation from the initial configuration) can substantively change the observed GROD angle maps [10].

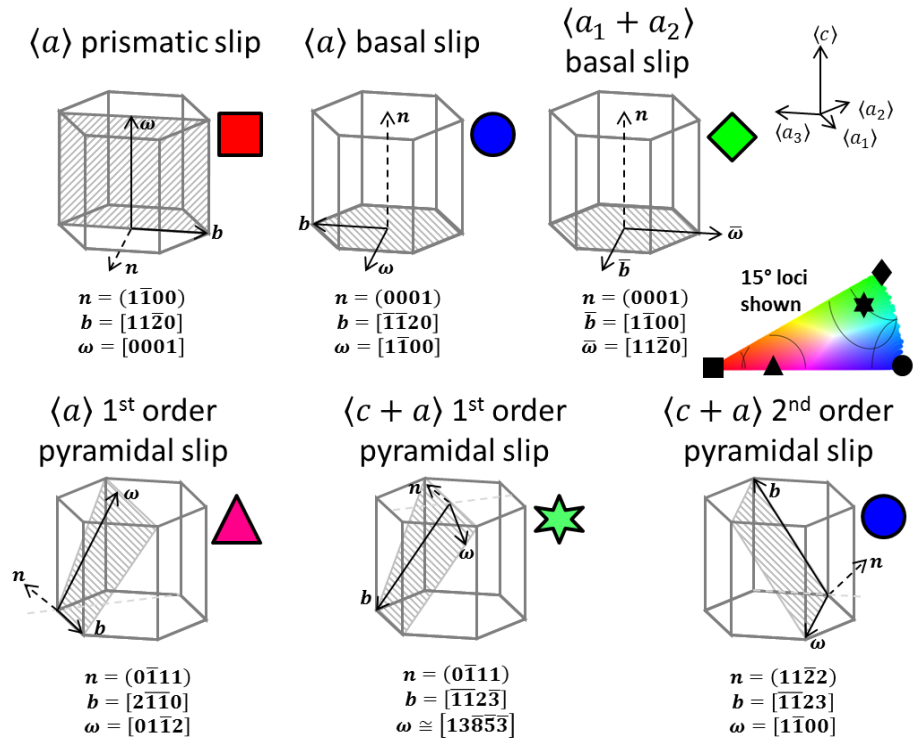

Figure 1: Hexagonal unit cells showing known slip systems in zirconium, showing slip plane normal **n**, slip direction **b**, and lattice rotation axis vector **ω**. **n**, **b** and **ω** form a right handed set. Vectors can be plotted and coloured according to an inverse pole figure (IPF). Lattice rotation axes for each slip system are plotted on the IPF triangle, and corresponding IPF colours are shown next to the unit cell drawings.

In Figure 1, we can observe that slip systems are reasonably well separated in the crystal reference frame based IPF colouring, with the exception of <a> basal and <c+a> slip on the 2$^{nd}$ order pyramidal slip plane. As previously measured [7], the CRSS of the <c+a> slip systems is much higher than <a> slip, and so it is more likely that <a> basal slip operates (though specific dislocation reactions and local stress states will activate <c+a> slip, especially as <a> type slip in insufficient to generate an arbitrary shape change in zirconium). Please note that we also consider the axis for double <a> type basal slip, which involves cooperative motion of two <a> basal dislocations with a collective Burgers vector that is perpendicular to an <a> type direction.

The axis of rotation can also be considered in the sample frame of reference. In monotonic and (constrained) uniaxial deformation, the expected axis of rotation is perpendicular to the principal strain axis. For the sake of the present work, we will describe this axis of rotation to be characteristic of 'homogeneous' deformation. The angle of deviation of the axis away from perpendicular indicates the degree in which the local region has a principal strain state that deviates from the external macroscopic boundary condition, or where a slip system with low critical resolved shear stress is favoured over a geometrically well-aligned slip system. This is likely driven by the local grain neighbourhood, where the tractions of the region surrounding the deforming volume are modulating the strain state to deviate significantly from the external boundary conditions.

If we assume that the Taylor axis reflects the local slip, as constraint is observed, then we can explore in which frame of reference the axis must be explored [18]. Formally, the Taylor rotation axis is constructed with respect to the undeformed configuration [22], and unless we perform an in-situ experiment or have an knowledge of the grain orientation with high precision we may not know this configuration, in particular when multiple slip systems contribute to the lattice rotation. As a proxy, an alternative rotation axis, such as the rotation axis with respect to the mean grain orientation or the local axis of the subgrain boundary has been used [18, 23-27].

In addition to this question, we can also note that quantitative analysis of the Taylor analysis requires use to have reasonable precision in the axis of rotation. However, as is well known [18, 28, 29] for axis-angle decomposition of a rotation matrix, the axis of rotation is difficult to analyse when the angle is small. Therefore, in the present work, we also assess the impact of this uncertainty through consideration of an uncertainty locus within our axis analysis (as shown by the circles within Figure 1).

In the present manuscript, we conduct Taylor axis analysis of deformed zirconium. We perform an *in-situ* monotonic and (macroscopically) uniaxial deformation with EBSD mapping at 0% strain and 12% strain. Using TrueEBSD [30], we spatially map our deformed EBSD map back into the undeformed spatial configuration and this enables us to assess the Taylor rotation axis in more detail.

## Method

Tensile specimens were cut from commercially pure zirconium purchased from Goodfellow. This rolled sheet had a typical 'split' basal ND texture, with the <c> axis of many grains pointing within 30° of the sheet normal direction. The samples were cut with along the plate rolling direction, to maximise deformation via slip and reduce the likelihood of twinning. At 12% uniaxial plastic strain, no twinning was observed. The geometry was optimised for *in-situ* testing in a Gatan Microtest 2000E. The gauge was 2 mm long (X – along RD), the sheet had a thickness of 1.71 mm (after polishing, Z along ND), and a gauge length of 1.33 mm (Y – along TD). The tensile test was performed using displacement control and yield was observed at 840 N and the test was stopped at 1024 N. These correspond to a macroscopic applied stress of 370 MPa and 450 MPa. The gauge length was measured after the sample was unloaded and it had deformed to 12% total plastic strain.

All orientation maps were post-processed (using True EBSD) to overlay onto a grain contrast image (FF-ARGUS) taken before deformation. Small changes in grain shape from deformation meant that the image registration does not fully match the boundaries, most boundaries have moved by a few pixels apart before and after deformation.

The EBSD data was analysed using custom codes, based upon the orientation conventions described in Britton et al. [31]. Two reference orientations were used for the GROD maps:

1. Undeformed reference – these maps compare the final orientations with the average grain orientation from undeformed grain at the same location, with orientation maps registered to the undeformed FF-ARGUS image. The undeformed grains have minimal internal misorientation, so use of the mean orientation is principally to reduce noise and enable comparison of location where there was no reasonable EBSD solution in the undeformed map.
2. Deformed reference – these maps compare the final orientations with the mean orientation within the same grain. The mean orientation was calculated firstly by trimming off 10% (outlier) orientations and then approximated by calculation of the arithmetic mean of quaternion components to minimise Euclidean distance in quaternion space [32], and when needed the quaternion can normalised to unit length. We include comparison with the average grain orientation in the deformed reference configuration for exploration of whether the deformed configuration (in absence of knowledge of the undeformed configuration) can be informative.

We present the GROD angle maps, where the angle indicates how much lattice rotation has can be expected (if the dislocation slip consists of constrained deformation). The second purpose of the angular maps is that we can use these to consider the uncertainty of the axis maps. As we have commented, it is well known that the axis of misorientation is more uncertain if the angle is small[1]. We will consider this uncertainty using loci of uncertainty within the IPF colour keys (see the dashed lines within Figure 1).

The GROD axis maps are presented in two frames of reference:

1. The crystal frame maps, using inverse pole figure colouring, can be used to identify the most likely active slip system. The colouring can be compared against the expected colouring of the Taylor axes for each slip system, shown in Figure 1 (subject to measurement uncertainty in EBSD misorientations). The colouring of the axis of lattice rotation may not correspond to any single slip system (within measurement uncertainty) and this is most likely due to the activation multiple slip systems activated during deformation and the rotation axis here is a path dependent composite of all the activated dislocation types.

2. The sample frame of reference. Here we choose to colour the maps by considering the angle between the rotation axis and the loading axis. If the axis of lattice rotation is perpendicular to the applied load, then the deformation (broadly) conforms to the macroscopic boundary condition. However, all locations within the deforming volume have no knowledge of the external boundary condition, and instead they are deforming subject to the local boundary conditions imposed their immediate neighbourhood. For anisotropic materials this results in strong neighbourhood effects [33] and therefore requires care when using the (macroscopic) Schmid Factor, such as required in a Sachs model. In our analysis, we can see this local neighbourhood effect when the rotation axis deviates strongly from 90° as expected from the macroscopic applied load, and we call these heterogeneously deforming regions.

---

[1] Consider flying around the world using great circles. The axis of rotation is a normal vector to the plane containing the great circle, and this vector goes through the centre of the earth. It is more difficult to describe the axis of rotation for a short flight from London to Paris as compared to flying from Beijing to San Francisco.

# Results

## 1.1.1 IPF orientation maps

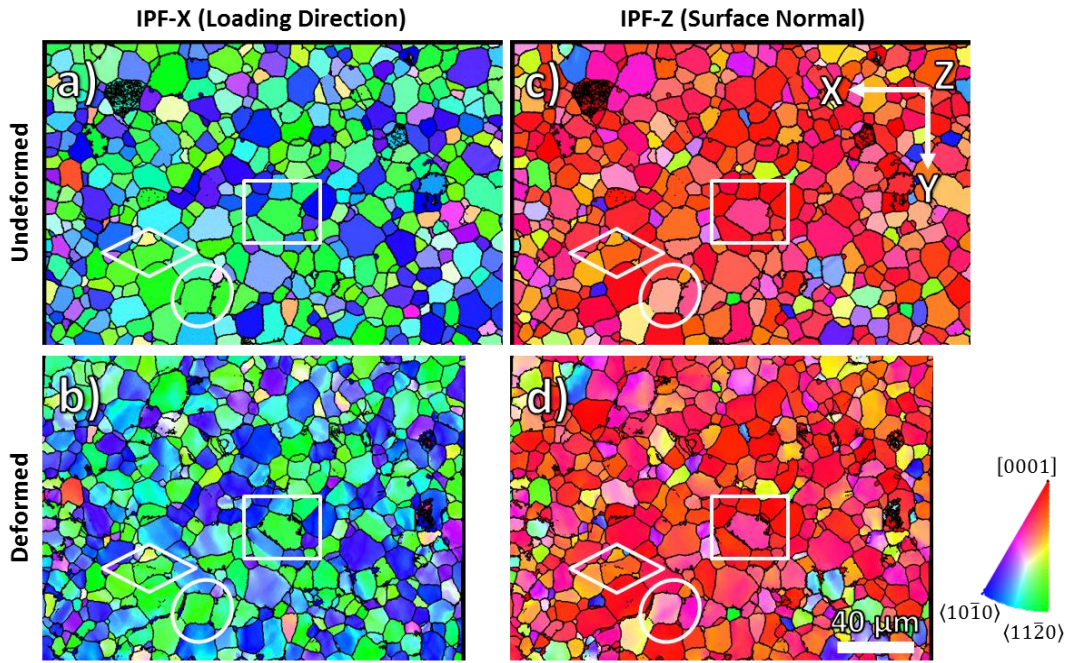

Figure 2: EBSD orientation maps of the tensile sample before and after deformation, in inverse pole figure colouring. The field of view is the same in both samples. The loading axis is parallel to X. The <c> axes of most grains are oriented along Z. a) IPF-X before deformation, b) IPF-X after deformation, c) IPF-Z before deformation, d) IPF-Z after deformation, with spatial remapping back to the undeformed configuration.

Figure 2 shows the crystal directions along the loading axis before and after deformation. As the deformed map is spatially registered with respect to the undeformed configuration there is shortening of these maps.

The undeformed EBSD map shows a typical rolled Zr-sheet texture where the majority of grains have <c> axes pointing towards the surface normal and the prismatic grains are pointed along the loading direction (Figure 2A and C). The intragranular misorientations before loading is small, where the modal GROD angle = 0.3° and this is limited by the conventional Hough-based EBSD angular resolution.

The deformed EBSD maps (Figure 2B and D) show that there is negligible twinning after deformation. The lattice orientations have changed. We can see orientation gradients as colour gradients in the IPF maps.

In these maps, and subsequent analysis, we have labelled three regions with a square, diamond, and ellipse. These will be analysed in detail later.

## 1.1.2 GROD angle maps

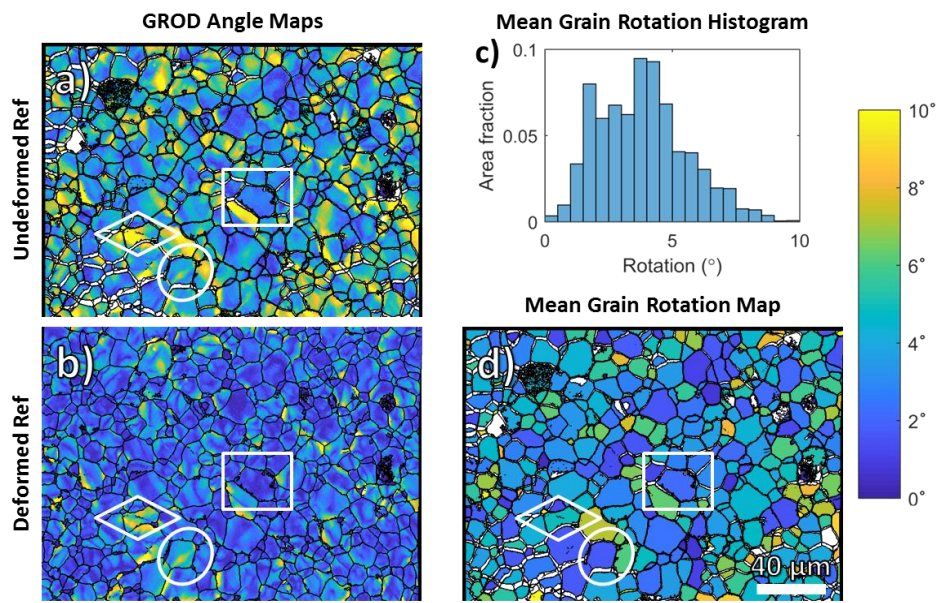

Figure 3: a) GROD disorientation angle map w.r.t. reference orientations in the undeformed configuration. b) GROD disorientation angle map w.r.t. reference orientations in the deformed configuration. c) Change in mean grain orientations from deformation, plotted as a frequency distribution of the disorientation angle. d) Change in mean grain orientations from deformation, plotted as a grain map.

Figure 3 shows GROD angle maps of the deformed structure, using reference orientations from deformed and undeformed grains. Typically, the change in lattice rotation is less than 10° for 12% average plastic strain.

Comparing the deformed orientations with the initial grain orientations (Figure 3A), we can see that slip within grains is heterogenous. There are regions where the lattice rotation is highest towards grain boundaries (e.g. the grain in the bottom left of the square), and in other grains there are bands of high and low lattice rotation that extend across the grain interior from one boundary to another, which are signatures that the grain orientation is fragmenting into two or more sub-grains.

The lattice rotation map comparing the orientation of each point with the mean orientation all in the deformed configuration (Figure 3B) shows that the magnitude of the lattice rotation is apparently much smaller. Heterogenous deformation remains visible, but the structures observed within can look substantively different to the total amount of point-based lattice rotation (Figure 3A). This is most noticeable in grains where the mean grain orientation has rotated substantially (Figure 3D) between the undeformed and deformed configuration, such that the mean grain orientation is no longer similar to the reference orientation of the deformed reference GROD map (Figure 3B).

The mean grain rotation map (Figure 3D) shows the change in the average orientation of each grain between the deformed and undeformed configuration. We see that this map is heterogenous, and there is a distribution of mean grain rotation (Figure 3C).

### 1.1.3 GROD rotation axes

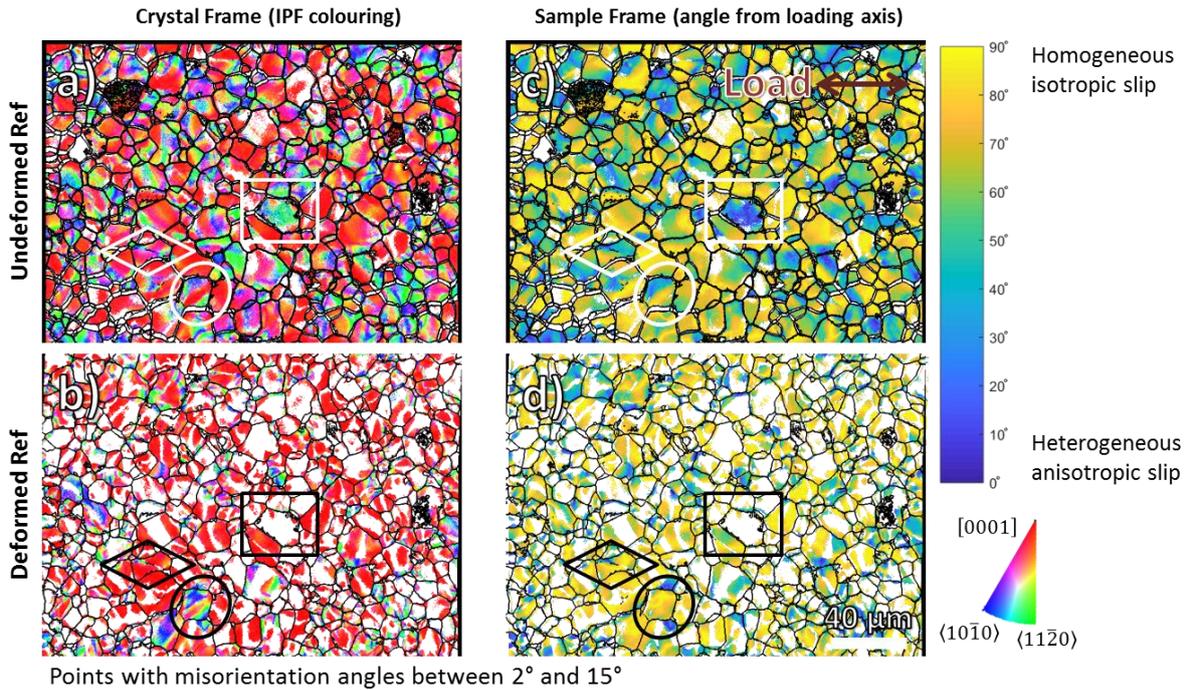

Figure 4: GROD disorientation axis maps, w.r.t. reference orientations in undeformed (a, b) and deformed (c, d) configurations. Points with corresponding GROD angle < 2° are removed due to high uncertainty in the GROD axis measurement. a) and b) show the axis as HCP crystal directions in inverse pole figure colouring. c) and d) show the axis in sample frame, plotted as the angle between the loading axis (horizontal) and GROD axis.

In Figure 4, we consider the GROD disorientation axis presented using the crystal frame (a and b) and the same frame (c and d). We have made points where the GROD angle is less than 2° white, due to the substantial uncertainty of the axis. We have also made points with GROD angles great than 15° white, as these are usually points near grain boundaries where the spatial registration is imperfect and the two orientations compared are from different grains as the local change in grain shape does not fit our simple model used in the TrueEBSD registration. Maps without these regions removed can be found as the Supplementary Figure 1.

The GROD disorientation axis map (Figure 4A and C) using the undeformed lattice orientation as reference show the most useful and representative understanding of the slip activity. When exploring the maps in of disorientation axis in the crystal frame (Figure 4A), broadly, we see that a substantive area fraction of this map is red indicating that the dominant slip system might be prismatic, however we note that the 'red coloured' faction of the IPF colour key covers a very large angular variation. The colour key in Figure 1 indicates that red is prismatic slip and 'off red' is more likely to be the activation of <a> pyramidal slip. We can also observe that there are substantive areas of the map which are coloured blue, which corresponds to activation of <a> basal slip or <c+a> pyramidal slip; and green, which corresponds to activation of double <a> basal slip (i.e. the cooperative motion of two <a> dislocations within the basal slip system).

Exploration of the disorientation axis with respect to the loading axis (Figure 4C) shows that the majority of grains tend to deform with the lattice rotation perpendicular, indicating homogenous slip. However, there are individual grains (e.g. inside the square) where the grain deforms heterogeneously, most likely according to the local neighbourhood. While this grain has a similar orientation to its neighbours (see Figure 2), the two-point correlations of the surface neighbourhood (and the subsurface neighbourhood, which is not explored here) seem to produce a strong neighbour effect. This additional results in a change in the locally activated slip system, where this grain seems to deform according to double basal slip (green in Figure 4A). In addition to these whole grain disparities, we also see the presence of bands that extend across more than one grain and are connected at the grain boundary. We could imagine that these regions of heterogenous slip are strongly controlled by the formation of intergranular slip bands. We can suppose that where the normal principal stress direction is not along the loading direction (Type 1 stress), the stress state is controlled by deforming grain neighbourhood (Type 2 stress). This can result in activation of slip with a 'poorly' orientated slip system if the global Schmid factor were to be considered.

The GROD disorientation axis map using the deformed lattice (Figure 4B and D) as reference show many more points which are uncertain, as the total angular deviation is much smaller (see Figure 3). Broadly we can see similar trends to consideration of the disorientation axis on average across the map, however the precise details of each point within the map may be substantively different. We can focus on these deviations by considering the specific highlighted grains.

In the rectangle grain we observe heterogeneous slip using undeformed reference GROD, but this is thresholded out of the deformed ref GROD data due to low GROD angle. This highlights that using the deformed reference results in loss of information due to uncertainty in the result.

For the diamond grain, the undeformed reference GROD axis map shows that top right half of the grain is rotating differently to the bottom left half and this is completely missed when exploring on the deformed reference GROD axis map.

For the ellipse grain, the GROD axis patterning is relatively more similar when comparing deformed and undeformed GROD axis maps, as the mean orientation changed by <2°.

## Discussion

Analysis of the change in lattice orientation via GROD analysis is a popular method of trying to understand the characteristic signatures of plastic deformation, particularly with the ease of access to this information from EBSD (and other orientation mapping techniques). Our repeat analysis, with tracking of the orientations from the undeformed to deformed configurations, was motivated by a previous study where we tried to track slip activity in shocked samples, which we tried to publish but it was subject to critique in peer review due to the uncertainty in the reference grain orientation. There are multiple studies in the literature where this analysis has been tried, but our findings here emphasise that we can misinterpret the axis if the mean grain orientation is used as the reference point (consistent with the discussion of [34]). From our work here, we note that small (1-9°) change in mean grain orientation leads to large change in patterning of GROD axis, and this is reflected in both sample and crystal frame. This complicates like-with-like analysis, as without a sufficient crystal plasticity-based understanding of the deforming state, analysis of the GROD axis within the deformed configuration (even for simple uniaxial tension) has limited value. Furthermore, the intragranular variations in GROD angle are also different between undeformed and deformed reference, as the sign and sense of the rotation can change. However, the GROD angle is an unsigned scalar value which is a simple extraction from the rotation tensor. It is not possible to simply add or subtract the mean rotation angle to get back the patterning (see the second Supplementary Figure).

Thus far, we have performed quantitative analysis of the change in lattice orientation but we have only provided qualitative assessment. Here we consider the area fraction of the maps and categorise slip activity. For each point within the map, we identify membership of the different regions within the regions identified in the IPF key (Figure 1). We perform this categorisation (Figure 5) through use of the two frames of reference, and different threshold values for our analysis.

In all cases, we observe that a significant fraction of the map (>50%) is described by multiple slip (i.e. the points are too far from the characteristic axis locations). We see that there are substantive regions (>3%) which can be described by <c+a> pyramidal 1 slip. While this slip system has a higher critical resolved shear stress than any of the <a> slip systems, it allows a substantially different change of shape of the unit cell. We observe a significant map area fraction (>10%) which can be described by the <a> pyramidal rotation axis, for single slip. We note also that if multiple slip was occurring, this axis could be a (weighted) linear combination of <a> basal and <a> prismatic slip.

The area fraction of <a> basal and $<a_1+a_2>$ basal slip is substantive, especially for the most reasonable analysis (undeformed, 15° max axis deviation, 2° min GROD angle). The presence of <a> basal slip is not disputed in these alloys, however double basal slip is considered more infrequently. Exploration within the GROD axis maps indicates that double and single <a> basal slip are often linked (e.g. in the grain within the centre of the ellipse highlighting and the grain in the centre of the rectangle, in Figure 4). At the length scale explored here, we would not require the double slip mechanism to happen on the same slip plane (indeed we do not know this information) and we could imagine that there could be cross slip involved. These regions of where <a> basal slip occur, for this crystallographic texture and loading configuration, tend to be regions with significant heterogeneous slip (see the GROD axis presented in the sample frame within Figure 4C).

Finally, in all cases we see a substantive area fraction of the map deforming with <a> prismatic slip. We note that our analysis in Figure 5 only considers the area fraction for single slip mechanisms, or the relaxed case for secondary slip activity (where the fraction of <a> prismatic slip increases). We could imagine that a significant amount of the multiple slip regions within the maps may be carried by <a> prismatic slip, acting together with another slip system, as the GROD axis map in Figure 4A is dominated by red colouring.

Comparison of the different thresholds and analysis methods shown in Figure 5 demonstrate that the GROD reference frame is important for determining activated slip systems.

Using deformed orientations as a GROD reference decreases the average GROD angle (Figure 2), and therefore the area fraction with low GROD axis uncertainty also decreases. This is evident in the rectangle grain in Figure 4, where the GROD axis is only accessible using reference orientation from the undeformed grain.

The area ratios between activated single slip systems is not sensitive to GROD segmentation or threshold angles (compare A(i), B(i), C(i); and A(ii), B(ii), C(ii)). This suggests that the 2° threshold angle for the GROD axis is adequate and any angular uncertainty is not producing misleading results.

However, if we force a tighter constraint on the labelling of each slip system, the regions ascribed to multiple slip increase substantively (as they are no longer ascribed to single slip).

The area ratio of active single slip systems changes significantly if the GROD reference is taken from the deformed grain orientations, instead of undeformed grain orientations. This is particularly evident for the area fraction of regions that are ascribed to <a> prismatic slip activity, where using a deformed reference GROD axis overestimates the area dominated by <a> prismatic slip.

This hints that it would be insightful to perform a repeat study of the same region within a sample with multiple EBSD maps may enable us to understand the progressive activity of these slip systems (and we could imagine including DIC and/or HR-EBSD based GND measurements [35]). Furthermore, this would be enhanced through analysis of the slip vector field from Heaviside [36] or standard digital image correlation [4].

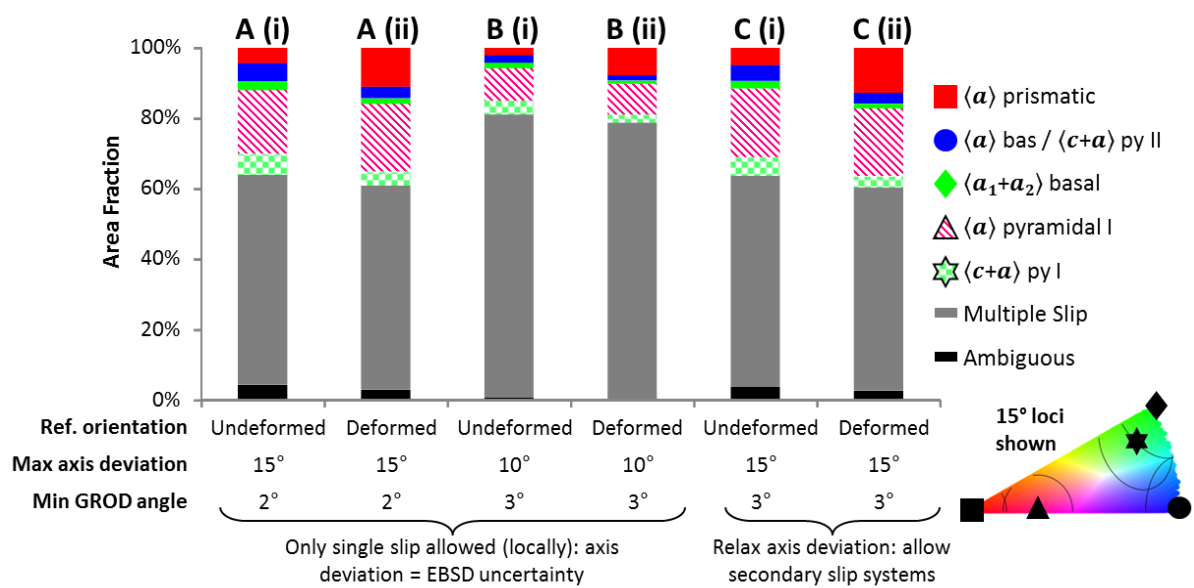

Figure 5: Area fractions of single slip systems measured from GROD axis in crystal frame, using either undeformed (i) or deformed (ii) reference orientations. Area fractions of single slip activation for each slip system are shown in A and B, corresponding to two axis measurement uncertainty values ('Max axis deviation'), which is a function of the minimum allowed GROD angle ('Min GROD angle'). In C, the axis deviation threshold is relaxed beyond the EBSD measurement uncertainty, to allow some secondary slip activation.

A final consideration when exploring this approach is that while we perform the analysis assuming that the material is constrained, for interpretation of the constrained lattice rotation via the Taylor axis, we are exploring the surface of polycrystal. Here, a change in surface roughness due to out of plane deformation, and the limited surface constraint (i.e. the plane stress configuration) may impact our interpretation, and we expect that a half-plane stress-based crystal plasticity analysis may be supplement our understanding of these contributions.

## Summary

Repeat mapping lattice orientations for the same area and overlap of the orientation data has enabled us to analyse the likely active slip systems. This analysis provides different view of the slip activity, which would normally only be accessible from post-mortem TEM analysis of the dislocation debris, analysis of in-plane shear from digital image correlation, or out of plane shear from slip-trace analysis. In assuming constraint due to the neighbourhood of the grains, we can track the rotation axis. In zirconium, the slip systems are well spaced across the orientation triangle and therefore it is possible to identify the most likely slip systems that can be used to describe the accumulated lattice rotation. For commercially pure zirconium deformed in tension, we observe significant lattice rotation that can be described by <a> pyramidal slip. Furthermore, while we observe that the macroscopic boundary condition provides an overall homogenous deformation field, we note that there are specific grains which deform heterogeneously, and this is likely due to the constraint of the grain neighbourhood. We imagine that this information complements field-based analysis of crystal plasticity in polycrystalline aggregates.

## Acknowledgements
VT and TBB acknowledge funding from Imperial College London and HEIF through the Proof of Concept scheme and EPSRC EP/ K034332/1 (HexMat). TBB acknowledges funding of his Research Fellowship from the Royal Academy of Engineering and funding from EPSRC EP/S01702X/1 (MIDAS). Electron microscopy was performed within the Harvey Flower Electron Microscopy Suite at Imperial College London and the Quanta was purchased within the Shell AIMS UTC.

## Data Statement
Data will be released to Zenodo upon paper acceptance.

## Author Contributions
VT and TBB designed the experiments, following initial discussions with EW. VT conducted the experiments and wrote the Matlab analysis code and drafted the initial outline of the paper. TBB wrote the paper. All authors contributed to the final manuscript.

# Supplementary Figures

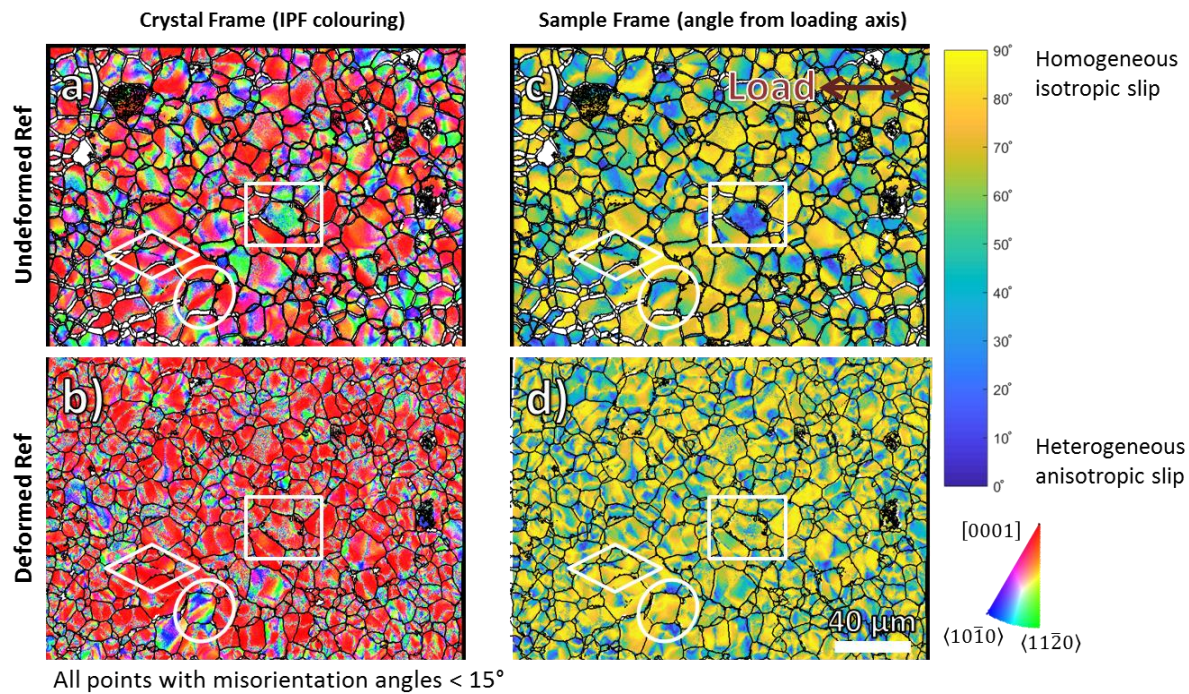

Supplementary Figure 1: GROD disorientation axis maps, w.r.t. reference orientations in undeformed [a, b] and deformed [c, d] configurations. Points with high measurement uncertainty are not removed. a) and b) show the axis as HCP crystal directions in inverse pole figure colouring. c) and d) show the axis in sample frame, plotted as the angle between the loading axis (horizontal) and GROD axis.